\documentclass[nofootinbib,amsfonts,preprint,aps]{revtex4}
\begin{document}

\title{Event ontology in quantum mechanics and downward causation}

\author{Rodolfo Gambini}
\address{Instituto de F\'{\i}sica, Facultad de Ciencias, 
Igu\'a 4225, esq. Mataojo, 11400 Montevideo, Uruguay}

\author{Jorge Pullin}
\address{
Department of Physics and Astronomy, Louisiana State University,
Baton Rouge, LA 70803-4001\\
Tel/Fax: +1 225 5780464, Email: pullin@lsu.edu}

\begin{abstract}
We show that several interpretations of quantum mechanics admit an
ontology of objects and events. This ontology reduces the breach
between mind and matter. When humans act, their actions do not appear
explainable in mechanical terms but through mental activity: motives,
desires or needs that propel them to action. These are examples of
what in the last few decades have come to be called ``downward
causation''. Basically, downward causation is present when the
disposition of the whole to behave in a certain way cannot be predicted
from the dispositions of the parts. The event ontology of quantum
mechanics allow us to show that systems in entangled states present
emergent new properties and downward causation.
\end{abstract}
\maketitle

\section{Introduction: The derivation of an ontology from physics}

Classical physics has inspired an ontology that that has had 
profound implications in the origins of modern philosophy. In this
ontology one assumes certain robust associations of properties and
objects. These associations have certain stability with respect to
time and are independent of the specific sequence of observations.
For instance sufficiently frequent measurements of the position of a
particle give similar values and if one decides to measure other
attributes between two measurements of positions the result remain
unchanged. These hypotheses, valid in any classical ontology, led to
Hume's conception of the world, so ingrained in contemporary
philosophy. We will return to them later on.

To derive an ontology from the quantum theory has not been easy given
its interpretational difficulties together with its conflict with the
classical ideas. In spite of initial efforts by the leading founders
of the field, dealing with the problem with an operational approach,
questions have lingered.  Within the Copenhagen Interpretation quantum
mechanics is introduced in terms of measurements.  Quantum systems do
not have definite properties till they are measured. Between
measurements their evolution is deterministic, as described by
Schr\"odinger's equation, but in the measurements one has probabilistic
behavior. In the standard presentation, one also has to assume that in
the measurement processes the states change in a way not described by
the Schr\"odinger equation. The difficulties to associate attributes to
a quantum system independent of it being measured or not are in the
core of the ontological problems presented by the quantum theory.

Perhaps the common characteristic between the different proponents of
the Copenhagen Interpretation is precisely that facing certain
ontological problems they retreat and stop any attempt to analyze the
situation in fully realistic terms. As we have seen, this point of
view appears very problematic to all those like Einstein,
Schr\"odinger or Bell, who refuse to take instrumentalist
positions. The acceptance without criticism of the Copenhagen
Interpretation in its various versions has not favored the development of
new ideas and questions and delayed the understanding of the most
revolutionary aspects of quantum physics. It was only due to the
immense intellectual stature of certain dissenters, as the ones we
just mentioned, that these problems remained in the spotlight and
allowed advances as those of Bell and the subsequent discovery of
quantum decoherence.  In the last decades this situation has changed
because the investigation of topics of great conceptual importance,
like those having to do with entanglement and decoherence, have opened
new perspectives with remarkable potential practical applications like
quantum cryptography or quantum computation. The advances have allowed
to verify that many macroscopic systems have quantum behavior and the
conviction that the quantum theory has universal validity has been
reinforced. It has also become apparent that the different physical
aspects related to the measurement problem like the interaction with
the environment or the use of physical clocks are perfectly analyzable
with the methods of quantum mechanics and should not be excluded at
the time of seeking a more complete understanding of the problem.

\subsection{The ontology of classical physics}

For centuries the principles of classical mechanics as formulated by
Newton and developed by Lagrange and Laplace were implicitly
considered as the basis and foundation of the scientific conception of
the Universe. The expectation was that the other sciences would
eventually be reduced and explained in mechanical terms.  Even though
this goal was never achieved many areas of knowledge adopted a general
mechanistic world-view. The mechanistic paradigm was superseded by
relativity and quantum mechanics but it has been extremely prevalent
until our days because of its simplicity and apparent consistency.  The
mechanistic paradigm is simple: matter is composed by elementary
components ---particles--- which are not altered when they combine to give
rise to complex structures. Classical mechanics identifies the world as
a succession of instantaneous configurations of systems of material
points that occupy successive positions in the mathematical space of
Euclidean geometry.

In this context, different phenomena produced by a system result from
the different configurations that its component particles take.  Given
the laws of force, the motion obeys a deterministic evolution.  Nothing
new may occur in a classical system that is not determined by its
initial configuration. In the classical world there is causal closure,
every event is the consequence of preceding events without any freedom
for novelty.  The elements of the classical world are matter, the
absolute space and time in which that matter moves, and the laws of
force that govern movement. No other independent categories of being,
such as mind, feelings or purpose are acknowledged.  Cartesian dualism
includes the mental aspects in terms of a new substance, being any
possibility of interaction between the mental and the physical
basically impossible to explain without ad hoc assumptions.

The discovery of an independent form of matter as the classical fields
---for instance the electromagnetic one--- did not change the basic
foundations of the classical ontology. In particular classical fields
have well defined attributes that can be measured at any time and the
theory still is determinist. The notion of separability that is at the
basis of Hume's doctrine of supervenience \cite{maudlin} is still
perfectly justified within the context of classical physics including
fields. It establishes that ``The complete physical state is
determined by (supervenes on) the intrinsic physical state of each
space-time point (or each point like object) and the spatio-temporal
relations between this points.''  In other words separability
establishes that the total state of the Universe is determined by the
states of its localized parts. As it was noticed by Teller
\cite{teller} and stressed by Maudlin \cite{maudlin} this notion of
separability no longer applies at the quantum mechanical level.

\subsection{Quantum mechanics and the crisis in the ontology of classical
  physics}

In the standard interpretation of quantum mechanics, the wave function
is a representation of all our probabilistic knowledge about outcomes of
possible measurements and as such is devoid of any ontological
content: As Busch \cite{classicalquantumontologybush} puts it:
``In other words in the standard interpretation, the formalism of
quantum mechanics or the quantum algorithm does not reflect a well
defined underlying reality, but rather it constitutes only knowledge
about the statistics of observed results.''

The classical concepts are put in doubt by this interpretation but are
not substituted by better, more suitable concepts.  Faye, in The
Stanford Encyclopedia of Philosophy
\cite{copenhageninterpretationfaye} summarizes Bohr's point of view as
follows:

\begin{itemize}

\item ``The interpretation of a physical theory has to rely on an
experimental practice.  

\item The experimental practice presupposes a
certain pre-scientific practice of description, which establishes the
norm for experimental measurement apparatus, and consequently what
counts as scientific experience.  \ldots

\item This pre-scientific experience is
grasped in terms of common categories like thing's position and change
of position, duration and change of duration, and the relation of cause
and effect, terms and principles that are now parts of our common
language.  

\item  These common categories yield the preconditions for
objective knowledge, and any description of nature has to use these
concepts to be objective.  
\item  The concepts of classical physics are
merely exact specifications of the above categories.  
\item  The classical
concepts... are therefore necessary in any description of the physical
experience in order to understand what we are doing and to be able to
communicate our results to others, in particular in the description of
quantum phenomena as they present themselves in experiments; ...''
\end{itemize}

But the consistency of the classical description is in this context in doubt
because the line of separation between the quantum object and the
measuring device is not the one between macroscopic instruments and
microscopic objects. In fact, as Bohr himself pointed out, parts of the
measuring device need sometimes to be treated in quantum mechanical
terms in order to have a consistent description of the measurement
processes.

\section{An ontology of states, objects  and events for quantum mechanics}

Attempts to base an ontology on events have a long history that was
reinforced by relativity and the quantum theory. In relativistic
physics events are considered points in space-time. The events of the
relativistic universe combine into partially ordered sets. An event B
is to the future of another event A if it belongs to its future region
defined by the light cone with vertex in A. In that case A can
influence B. If B is outside the future light cone of A then it is
causally disconnected with A since the maximum speed of propagation of
a signal is the speed of light.  The formalism of quantum mechanics
makes reference to primitive concepts like system, state, events and
the properties that characterize them. The use of these concepts
suggests that the theory should admit an ontology of objects
(understood as systems in given states) and
events.  The program of accounting for physical reality in terms of
events has a long and noble tradition that goes back to Russell, who
stated that \cite{theanalysisofmatter} ``the enduring thing or object
of common sense and the old physics must be interpreted as a
world-line, a causally related sequence of events, and ... it is
events and not substances that we perceive.''  To put it differently,
for Russell and object is nothing more than a set of events that are
causally connected.  Although we consider this point of view a step in
the right direction, we think it is incomplete for a foundation of
physical reality based on events, particularly in the light of quantum
mechanics.

A quantum system is described by a Hilbert space that represents the
set of its possible states and the events that may occur in the
system. Based on this ontology, objects and events can be considered
the building blocks of reality. Objects will be represented in the
quantum formalism by systems in certain states. In an event
interpretation like the ones we are considering, events are the actual
entities while states represent potentialities to produce events.

The basic idea of a measurement is the occurrence of a macroscopic
phenomenon, that is of something capable of reaching perception. Thus,
as noticed by Omn\`es, the measurement of a property of a microscopic
object implies making it generate a phenomenon, in other terms produce
an event. The process of detection of photons by dissociation of
silver bromide in a photographic plate leading to a cascade effect
that produces the accumulation of millions of atoms of silver is an
example of the production of an event. The appearance of a dot in the
photographic plate is an example of a macroscopic event that constitutes the world accessible to our senses \cite{omnes}.  The dot and its properties have,
like any property, a mathematical counterpart in the formalism of
quantum mechanics corresponding to projectors in the Hilbert space of
the detecting plate.  We are thinking in this kind of events as the
building blocks of the apparent reality.  Both the event ---the
appearance of a dot--- and its properties are characterized by
projectors \cite{axiomatic}.  We shall call essential property the
projector that completely characterizes the event and denote it by
$P_E$ that is a projector on the Hilbert space of the photographic
plate. Let us call $P_1,...P_n$ the projectors in the Hilbert space of
the plate that characterize the properties of the dot. As we have
shown in \cite{axiomatic} the properties of an event satisfy
$P_iP_E=P_E$.

States describe the potentialities or dispositions of the systems for the
production of certain events. The formalism of quantum mechanics
associates a projector to each property or event.
The element hydrogen is a quantum system.  A particular atom is a
system in a given state. It is an example of what we call object. It
is characterized by its disposition to produce events on other
systems: for instance the emission of a photon that produce a click in
a photodetector. Note that for Russell an atom cannot be considered an
object as long as it does not interact yielding events, whereas our
definition naturally includes any microscopic object as an object
given that its disposition to produce events is always defined by
its state.

Concrete reality accessible to our senses is constituted by events
localized in space-time. That is, by certain entities that occupy a small
region of space-time.  As Whitehead \cite{whitehead1925} recognized:
``the event is the ultimate unit of natural occurrence.'' Events come
with associated properties. Quantum mechanics provides probabilities
for the occurrence of events and their properties.  When an event
happens, like in the case of the dot on a photographic plate in the
double slit experiment, typically many properties are actualized. For
instance, the dot may be darker on one side than the other, or may
have one of many possible shapes. The postulated association between
properties and objects typical of the classical physics is now
substituted by an association of properties with events. Objects
understood as systems in certain dispositional state do not have
properties until their are measured or produce events

There is only an exemption to this rule. One can in principle assign
some properties to pure states. These properties are the ones observed
during the preparation of the state.  But, contrary to what happens in
classical physics these associations are not independent of the
specific sequence of observations performed on the quantum
system. Thus one of the main postulates of classical ontologies as
Hume's one is no longer valid. This assumption, that was considered by
Einstein at the root of the possibility of doing science\footnote{in
  one of the letters to Max Born Einstein says: ``the concepts of
  physics relate to a real outside world, that is, ideas are
  established relating to things such as bodies, fields, etc., which
  claim a 'real existence' that is independent of the perceiving
  subject. ... It is further characteristic of these physical objects
  that they are thought of as arranged in a space-time continuum. An
  essential aspect of this arrangement of things in physics is that
  they lay claim, at a certain time, to an existence independent of
  one another, provided these objects 'are situated in different parts
  of space'. Unless one makes this kind of assumption about the
  independence of the existence (the 'being-thus') of objects which
  are far apart from one another in space which stems in the first
  place from everyday thinking - physical thinking in the familiar
  sense would not be possible.''}  and the derived notion of
separability we mentioned before do not apply to quantum physics
that nevertheless is a rigorously formulated and tested theory.

When one considers non-local systems like particles in entangled
states, whose components occupy different positions in space-time it is
not possible to speak of a state at a given time, since that is a
notion that depends on the Lorentz reference frame chosen. However, if
the state is defined by its disposition to produce events one can
rigorously show \cite{GaPo} that such disposition is uniquely defined
and the state in the Heisenberg picture only changes when events take
place. The disposition to produce events separated spatially in the
sense of relativity, that is not causally connected, is independent of
the temporal order that one assigns to such events. In fact, 
the assigned order is purely conventional since it depends on the
reference system used. The concept of states in quantum systems is
necessarily holistic in space-time \cite{maudlin}. Very far removed
from the notion of separability that Einstein considered mandatory in
order to do science.

The event ontology we have presented has the attractive feature of
reducing the breach that exists between the material and mental
worlds. As Russell \cite{mind} pointed out ``if we can construct a
theory for the physical world which makes its events continuous to
perception, we have improved the metaphysical status of physics''.
According to his view we need ``an interpretation of physics which
gives a due place to perceptions''. The ontology of events we are
proposing could provide this interpretation: events in the external
world are subject to a physical description while at least some events
in our brain could be directly accessible as perceptions. Both mental
events and physical events would admit the same mathematical
description in terms of projectors in a Hilbert space. The main
difference between both forms of events is the way we access to them:
a first person access for the mental and third person access for the
physical As noted by David Chalmers \cite{chalmersfirst}: ``The distinguishing mark of the
first-person view is the air of mystery which surrounds it. This
feeling of mysteriousness has led many people to dismiss the
first-person out of hand. ... But the first-person is not to be
dismissed so easily. It is indeed a glaring anomaly today, in the
heyday of the scientific world-view. If it was not for the direct
experience which all of us have of the first-person, it would seem a
ridiculous concept. But it throws up too many problems to be neatly
packaged away in the kind of third-person explanation which suffices
for everything else in the scientific world. Pity.''

Each primitive concept that is introduced in the axioms of quantum
mechanics is associated with a mathematical concept well known in
ordinary quantum mechanics, but one can only assign them a well
defined philosophical meaning if one has an interpretation of the
theory.  For example, quantum mechanical events could not be used as
the basis of a realistic ontology without a general criterion for the
production of events that is independent of measurements. On the other
hand, the concepts of state and system only acquire ontological value
when the events also have acquired it. It is important to remark that
having a realist interpretation of quantum mechanics not only allows
us to understand the measurement process; it also allows understanding
how a world with uniquely defined properties arises from a quantum
world of potentialities. Some of these interpretations will allow us
to derive an ontology where objects and events can be considered the
building blocks of reality.

\section{Interpretations that admit an event ontology}

In what follows we will discuss three interpretations that admit
event ontologies. They are the Many Worlds Interpretation, the Modal
Interpretation and Montevideo Interpretation. As we will see not all
of them lead to the same notion of event. As a consequence the
corresponding ontologies will exhibit minor variations depending on
what interpretation is considered.

\subsection{Events in the Many Worlds Interpretation}

Everett \cite{everett} 
addressed the measurement problem assuming that the wavefunction
describes the Universe as a whole, including the observers and that it
evolves continuously obeying the Schr\"odinger equation without any
discontinuity or collapse.

In order to explain our observations he assumed that the wave function
of an observer would, in effect, bifurcate at each interaction of the
observer with a superposed object. The universal wave function would
contain branches for every alternative result of the measuring
object. Each branch has its own copy of the observer, a copy that
perceives one of those alternatives as the outcome. Schr\"odinger
evolution ensures that once formed, the branches do not influence one
another. Thus, each branch embarks on a different future,
independently of the others.

In order to understand how the branches corresponding to different
measurement outcomes become independent and each turn out looking
classical one uses today the decoherence theory. For Everett, all
elements of a superposition (all ``branches'') are ``actual'', ``none any
more real than the rest.'' Summarizing: worlds ``are mutually 
dynamically isolated structures
instantiated within the quantum state, which are structurally and
dynamically quasi-classical'' \cite{Wallace}.  The existence of possible 
different ``worlds'', is established by decoherence theory.

The ontology of the Many Worlds Interpretation may be considered as a
monist perspective where states and events have the same nature. One
may consider that in this approach what we call states and events are
aspects of a fundamental entity which is the state of the complete
multiverse. Notice that the relevant kind of events in the Many Worlds
Interpretation corresponds to events associated with observable
phenomena that give rise to different ``mutually dynamically isolated
structures instantiated within the quantum state''.

But one needs to assume that for certain superpositions ---that
according to decoherence are very similar to statistical mixtures---,
the different states that compose the statistical mixture acquire
independent reality. The appearance of independent realities in the
branching process may be considered as the events. But strictly speaking 
one does not have statistical mixtures but
superpositions, and it is not clear how to assign any reality to these
superpositions, given the fact that different basis may be used to
define them. It is not our purpose to enter into a critical analysis of the
interpretations, however we remark that the ontology itself is
problematic in the Many Worlds Interpretation.

\subsection{Events in Modal Interpretations}

Van Fraassen \cite{bvf} proposed an alternative procedure to eliminate
the projection postulate from the quantum theory. His proposal relies
on the distinction between ``dynamical states'' and ``value states''
or ``actual-valued'' observables. The dynamical state is the usual
state of quantum mechanics: it determines which properties may the
system have in its future and their corresponding probabilities. The
value state represents the physical properties that the system
actually has at a given instant. The Modal Interpretation in its
different versions assumes that physical systems at all times possess
a number of well-defined physical properties, these properties can be
represented by the system's value state.  An essential feature of this
approach is therefore that a system may have a sharp value of an
observable even if the dynamical state is not an eigenstate of that
observable. What changes in the different versions of this
interpretation is how the actual valued properties are defined.

In the Kochen--Dieks (K-D) \cite{kochendieks} 
Modal Interpretation the biorthogonal
(Schmidt) decomposition of the pure quantum state of the system
selects the actual-valued observables.  In the Vermaas--Dieks (V-D)
\cite{vermaasdieks}
version the actual-valued observables are defined by the spectral
resolution of the system's reduced state, obtained by partial tracing.
Even though these proposals are well suited to describe ideal 
measurements they failed to describe imperfect measurements. In fact,
they do not select the right properties for the apparatus in the imperfect case 
(see Albert and Loewer \cite{allo}).

Castagnino and Lombardi have observed that the Hamiltonian of the
quantum system plays a decisive role in the property-assignment rule
that selects the observables whose possible values become actual.
Once $H$ is given, it is assumed that the actual-valued observables of
the system $S$ are $H$ and all the observables commuting with $H$ and
having, at least, the same symmetries as $H$.  This is the Modal
Hamiltonian Interpretation (MHI).  Independently of the particular
implementation of the Modal Interpretation adopted one can consider
that each time a property is instantiated in a system an event occurs.

As Lombardi and Dieks observe in \cite{SEP}: ``In modal
interpretations the event space on which the (preferred) probability
measure is defined is a space of possible events, among which only one
becomes actual. The fact that the actual event is not singled out by
these interpretations  is what makes them fundamentally
probabilistic. This aspect distinguishes modal interpretations from
many-worlds interpretations, where the ``probability measure'' is
defined on a space of events that are all actual. Nevertheless, this
does not mean that all modal interpretations agree about the
interpretation of probability.'' We share the MHI point of view that
adopt a possibilist conception, according to which possible events
{\em possibilia} constitute a basic ontological category (see Menzel
\cite{menzel}).  The probability measure is in this case seen as a
representation of an ontological propensity of a possible quantum
event to become actual \cite{lombardicastagnino2008}. The Modal
Interpretation does not assume that the state changes after a property
instantiates and it assumes that the evolution is always
unitary. Nevertheless it is assumed implicitly that after the
observation of a property the disposition of the system in its
subsequent evolution is the same as the one the system would have had
if the state had collapsed.  Therefore, the ontology of states and
events appear to be well suited to this interpretation. However, 
here the relevant kind of events corresponds to
instantiations of properties both of macroscopic and microscopic
systems and therefore they do not necessarily correspond to phenomena.

We have already expressed some reservations about the Many Worlds
ontology. In the case of Modal Interpretations, we also have
reservations which are related again with some implicit
approximations. In this case with the notion of ``elemental quantum
system'', that introduces an ambiguity in the definition of the
``actual valued observables''.  In fact, in order to have a well
defined Hamiltonian one needs to assume that $S$ is an ``elemental
quantum system'', that is a system which is in tensor product with the
rest of the Universe.  Strictly speaking, only the whole Universe may
be considered as an elemental quantum system, that is a system that do
not interact with its environment, and therefore the MHI is again -as
the Many Worlds Interpretation, based in an idealization. In this case
the idealization of an isolated system.

\subsection{Events in the Montevideo Interpretation of quantum
  mechanics}

The closest we have to an explanation for the measurement process
within the quantum theory is environmental decoherence. It is based on
the fact that when a quantum system interacts with an environment with
an enormous number of (microscopic) degrees of freedom, the state of
the quantum system suffers transitions that {\em almost} look like the
abrupt evolutions one needs to postulate in measurements. However,
even if they are difficult to detect, quantum superpositions are still
there and may be in principle observed.

Environmental decoherence is important because the measurement
processes involve interactions with macroscopic systems with many
degrees of freedom.  Interactions with the environment were neglected
for decades and the relevance of this effect was only recognized in
the 1980's

A new interpretation was recently proposed. The novelty of this
interpretation of quantum mechanics is the inclusion in the quantum
description of another factor up to now neglected. In the standard
Schr\"odinger description of the evolution, time is treated as a
classical external parameter, but time is actually measured by
physical clocks that obey quantum mechanical laws. Quantum
measurements of time have a limited precision. This limitation arises
from quantum fluctuations and gravitational time delay and has a
fundamental nature \cite{saleckeretal}.  This effect, first noticed
more than 50 years ago has been recently confirmed by many authors.
We have shown that a quantum mechanical treatment of time
\cite{datolo} combined with fundamental limitations of measurements
stemming from general relativity, lead to a modified Schr\"odinger
evolution that allows transitions between quantum superpositions and
statistical mixtures.

When one takes into account the limitations in measurement imposed by
quantum mechanics and gravity, the states resulting from decoherence
are indistinguishable from those produced by the measurement
postulate.  The ``almost'' of the standard approach to decoherence is
removed by fundamental limitations predicted by the theory itself and
the transitions from superpositions to statistical mixtures required
to explain measurements are deduced.  This in turn supplies an
objective criterion that says when and what events may occur. Events
occur when the state of a system resulting from a full quantum
mechanical evolution becomes a statistical mixture
\cite{axiomatic,review}. The transition of the state of the system
plus environment to a statistical mixture gives necessary and
sufficient conditions for the occurrence of events. Events are assumed
to occur as random choices of the system. They simultaneously lead to
the production of events and the state reduction. It is not assumed in
this interpretation that these choices are part of a process that the
theory describe as it is the case of the Ghirardi--Rimini--Weber
approach. It is an additional postulate that however needs no
reference to a classical world or a notion of measurement. 

The modified evolution induced by the use of a quantum time leads for
a quantum system coupled with its environment to a state that is a
statistical mixture.  This provides an objective criterion for the
occurrence of events and state reductions that establishes when events
and changes in the state may occur without disrupting the prediction
of the evolution equation. It is important to remark that within this
interpretation events always occur in systems that include a
macroscopic environment and therefore are macroscopic as the phenomena
considered by Omn\`es. In this sense this interpretation
keeps more similarities with the Many Worlds Interpretation than with
the Modal ones.

Up to now one only has a precise analysis of the complete process
leading to the statistical mixture for spinning particles. For other
systems one can prove that the state of the microsystem coupled to the
environment approaches exponentially to the statistical mixture. We
consider that this is enough to assume the indistinguishability. Given
the fact that the distinction between an evolution that includes
quantum time measurements or a quantum reduction would require an
exponentially growing number of individual measurements in order to
have the required statistics for distinguishing a non vanishing
exponentially small mean value from zero.  Limitations referring to
the existence of a finite number of physical resources in a finite
observable Universe would be enough to ensure
undecidability. \cite{Butter,review} However, this is a point that
needs further study to have a definite answer.

Summarizing, for all the interpretations that admit an event ontology
and for quantum systems interacting with a macroscopic environment
that has many degrees of freedom, events will be plentiful. They not
only occur on measuring devices, they occur around us all the time. 
Measurements are nothing but the assignment of quantitative properties
to events occurring in measuring devices.

\section{Emergence in terms of an event ontology}

Emergent phenomena are said to arise out of and be sustained by more
basic phenomena, while at the same time exerting a ``top-down'' control,
constraint, or some other sort of influence upon those very sustaining
processes. We are interested in strong emergence, its defining characteristics
are qualitative novelty and ontological non-reducibility.  The
notion of emergence we are proposing considers emergent
entities to be genuinely novel features of the world. By definition, one 
talks here of causal powers that cannot be explained in terms of the 
micro causal powers but arise from the existence of certain macro level 
entities.

Strong emergence seems to be particularly relevant: if someone attempts
to explain natural phenomena without denying the existence of mental
processes in physical terms she must demonstrate the viability of
emergence with downward causation. The central point which this
concept refers to is that higher level mental events have the ability
to influence the behavior of more basic levels.  This requires
philosophical compromises that are not easy to justify in terms of
an ontology based on classical physics.  In fact attempts of explanation
within a mechanistic view are problematic. 

There have been attempts Sperry \cite{sperry1987} to supplement
classical mechanics with ``configurational forces'' to account for
strong emergence and downward causation.  McLaughlin
\cite{mclaughlin1992}, explains this concept as follows, ``Consider
the doctrine that there are fundamental powers to influence motion
associated with types of structures of particles that compose certain
chemical, biological, and psychological kinds. Let us see what this
would imply in the framework of classical mechanics, for example. It
would imply that types of structures that compose certain special
science kinds can affect the acceleration of a particle in ways
unanticipated by laws concerning forces exerted by pairs of particles,
general laws of motion, and the spatial or spatio-temporal
arrangements of particles. In a framework of forces, the view implies
that there are what we may call ‘configurational forces’: fundamental
forces that can be exerted only by certain types of configurations of
particles, and not by any types of forces between pairs of particles
as the usual interacting forces of the fundamental systems as the one
exerted by charged particles.''

Let us examine these proposals in some detail. For that we need to
briefly discuss how systems of particles are described in classical
physics. In three dimensional space one needs three numbers in order
to characterize the position of a particle: its coordinates. One also
needs three numbers to characterize its velocity (one for its
magnitude and two for its orientation). Therefore six numbers tell us
all we need to know about the particle from a mechanical
perspective. With such information any property of the classical
particle can be determined and also its future behavior. If one has N
particles one needs 6N numbers. Such numbers can be thought of as
belonging in an abstract 6N dimensional mathematical space called
phase space. The evolution in time of a system of particles in such a
space is a curve in phase space. Any property of the complete system
at a given instant is determined by those 6N numbers. A way of
introducing configurational forces is considering a certain region of
phase space. The force acts if the point representing the state of the
system at a given instant of time lies within that region. Could such
a force account for downward causation? Notice that by referring to a
region of phase space, the force is therefore dependent on the
positions and velocities of all particles and not on pairs of
particles as are all the other fundamental forces of physics. Such a
force is not a combination of ordinary Newtonian forces nor results
from the action of gravity or electromagnetism. Obviously from the
point of view of ordinary physics the introduction of such kind of
forces, which depend on each system, is highly artificial. And since
they are not reducible to elementary forces one would need a new
ad-hoc force for each type of emergent system. This would therefore
preclude any scientific explanation of phenomena. This should require
a kind of ``intelligent design'' for any emergent system from a molecule
to a living being incompatible with any scientific
explanation. Attempts to understand emergence from classical physics
are destined to fail and have led those that followed this path to
believe, like Bedau does \cite{bedau} that ``strong emergence starts when
scientific explanation ends.'' 

We will show that certain quantum mechanical systems present precisely
this kind of top-down control. It is well known that in quantum theory
the physical state of a system of particles cannot always be reduced
to the state of its component particles, ``or to those [states] of its
parts together with their spatiotemporal relations, even when the
parts inhabit distinct regions of space.'' as noted by Maudlin
\cite{maudlin1}.  We want to emphasize here that this quantum
mechanical holism involves systems that have ontologically new
properties and present downward causation where macro-systems have
effects on their micro components \cite{uslewowicz}.  That is, the
basic tenet of strong emergentism is that at a certain level of
physical complexity novel properties appear that do not result from
the properties of the parts of the system or their relations and that
contribute causally to the world. That is, emergent properties have
new downward causal powers that are irreducible to the causal powers
of the properties of their underlying base. Ontological emergentism is
therefore typically committed to downward causation, that is, causation
from macroscopic levels to microscopic levels.  If we adopt Crane's
\cite{crane} 
terminology our position should be considered as non reductive
physicalism because it denies ontological reduction but admits
explanatory reduction in the sense that the upper level properties and
causal powers can be explained in quantum mechanical terms.

The traditional objections to emergence result from the explicit or
implicit use of ontological concepts based on classical physics and
are not tenable when assessed from a quantum mechanical ontology
of events.

\subsection{Ontologically new properties}

Let us start by showing that quantum systems may have ontologically
new properties. Quantum systems may have certain quantum states,
called entangled, that have well defined properties that neither
follow from the properties of parts, nor from relations among them. To
understand this statement better, let us review how entangled states
are defined and contrast them with systems in classical states.

In classical physics the state of a system of particles
$(\vec{r}_1,\vec{v}_1,\ldots,\vec{r}_N,\vec{v}_N)$  is simply the
union of the states of each particle. Its knowledge determine all the
properties of the system. For instance the energy
$E(\vec{r}_1,\vec{v}_1,\ldots,\vec{r}_N,\vec{v}_N)$. All the
properties of a classical system are functions of the properties of
its components.

In quantum mechanics things are very different. Most of the properties
of a system do not have well defined values until measured; for
instance the position of an electron is not well defined until a dot
is produced in the photographic plate and it is detected.  However any
quantum system in a pure state has some well defined properties.  For
instance, a spinning particle can take two possible values of its
component along an axis $\hat{z}$: up or down.  When one performs
repeated measurements on particles on a given state one observes dots
appearing in the upper region with certain probability an in the lower
with the complementary probability.  When the electron is in a state
that leads with certainty to a dot in the upper region of the detector
identified by $z>0$, one may say that it is in the state $\vert z,
{\rm up}\rangle$. In this case one may assign the property ``{ z up}''
to the state This is the only property that one can assign to this
state. The measurement of any other component will not lead to a
unique value: i.e. always up or always down. In general, the
properties of a pure state $|\psi>$ is always associated with a set of
projectors $P_1...P_N$ such that $P_i|\psi>=|\psi>$. It is only when
one knows with certainty what will be the behavior of the system in
certain state that one may assign it a property.  In fact, as we have
observed before, events have many well defined properties but
typically states do not have properties until they produce events.
One may assign properties to states only indirectly trough the
properties observed in some events produced during the preparation
process of the state.

Systems composed of several particles may also have states with some
properties with well-defined values. However, these properties may refer
to the system as a whole and, in these systems, there may not be any
property for the states of individual particles with well-defined
values. These composite systems are called entangled.
More in general, entangled systems are those that have properties with
well defined values than cannot be inferred from those of their
constituent parts.  As we will see in what follows, it might even be
the case that the constituent parts have no well defined properties
and yet the system as a whole does.  

Consider two electrons with spin in the z direction in a state
\begin{equation}
\vert \psi_0\rangle = 
\frac{1}{\sqrt{2}}\vert 1,z,{\rm up}\rangle\vert 2,z,{\rm down}\rangle
+
\frac{1}{\sqrt{2}}\vert 1,z,{\rm down}\rangle\vert 2,z,{\rm up}\rangle.
\end{equation}

Neither the state of particle 1 nor the one of particle 2 have well
defined properties. No matter what component of the spin of one of the
particles one measures, one has a probability $1/2$ of measuring up
and $1/2$ of obtaining down.
Even though each entangled electron do not have well defined
properties for their spin components the total system does. For
instance one can show that it has total spin $s=1$ in Planck units ,
and z component of the total spin $s_z =0$. It is only when the
observations made on particle 1 and 2 are compared that one can
discover the properties of the total system. One could also determine
these properties when the complete system is measured.
The constituents therefore now form an inseparable unit endowed with
properties without the individual systems having any property ---any
spin component--- with well defined value.

This holistic behavior is actually not an exception but is the generic
behavior of two quantum systems after an interaction. For instance, the
precise vibrational modes of a molecule depend on the entangled system
of electrons and nuclei.  Underlying this feature is the exponential
growth of states with the number of component particles in quantum
mechanics in contraposition with the linear growth in classical physics.
 Most of these states and their corresponding properties
---projectors--- would have never occurred in systems with independent
---non entangled--- components.

Ontological novelty manifests itself in the emergence of new
properties that do not result from properties of the parts. They arise
only when the composite structure is constituted. 

The emergent properties of such systems are crucial for explaining
 chemical or biological properties in physical terms. 
 For instance: the magnetic counterpart of the properties of
entangled spins work as a magnetic needle that is at the basis of
navigational skills of the European robin, a migratory bird able to
detect the direction and strength of the Earth magnetic field. A
``sixth'' sense known as magnetoreception \cite{mcfaddenalkhalili}.

The philosopher of science Paul Teller \cite{teller} was the first in
noticing that quantum phenomena show relations that do not stem from
non-relational properties of their relata, as is characteristic of the
classical description of the world.  Entangled systems present what
Teller calls: relational holism \cite{tellerbjp}. The emergence of new
properties of the whole in a quantum world where events and properties
play a fundamental role is a crucial manifestation of ontological
novelty. This conclusion follows provided one adopts a realist ontology.

Healey \cite{healey} has introduced the notion of 
Physical Property Holism that assumes that there are physical objects  ``not all of whose
qualitative intrinsic physical properties and relations supervene on qualitative intrinsic physical properties and relations in the
supervenience basis of their basic physical parts.'' He observed that the existence of  physical property 
holism in entangled systems
depends on the interpretation of quantum mechanics that one adopts. 
Our discussion was restricted to the ontology of events and
it is in this context that have proved that ontologically new properties arise.

\subsection{Downward causation}

A strong form of emergence also requires downward causation, namely,
the emergence of novel causal powers. Here a double goal arises:
to characterize such form of causality in physical terms and to show
that at least certain systems, like the quantum ones, exhibit downward
causation.

A notion of causality that is suitable to the ontology of states and
events has been developed by Chakravartty \cite{chakravarty}. He
founds his notion on what he calls "causal properties". As we have
associated the notion of property to projectors in the Hilbert space
we prefer to speak about causal powers.  Following Chakravartty we
will recognize a causal power for its capacity to confer dispositions
on the objects that have them to behave in certain ways when in the
presence or absence of other objects with causal powers of their
own. This dispositional idea of causality 
---originally due to Heisenberg---
is the one we have adopted
in our work for states: recall that quantum states characterizes
dispositions to produce events in their interaction with other
systems.

Causality is normally presented in terms of related events. In quantum
mechanics this vision is incomplete if the concept of state is not
included. Indeed, among the events that prepare a state and the ones
observed usually there is a period of time and the disposition to
produce events is not given just by the initial preparation. It is
indeed given by the state. The latter is defined in terms of initially
observed events and their time evolution. Changes in the state are
determined by its Hamiltonian evolution. Only the state at a time $t$
defines completely the causes that lead to the observation of events
in $t$. Quantum mechanics introduces a probabilistic notion of
causation that involves states as causes and events as effects. 

An object will present downward causation if the parts have some
behaviors that are dictated by the state of the whole and that cannot
be predicted from the knowledge of the state of the parts.  The
previous example of an entangled state shows that in quantum mechanics
there is state non-separability \cite{healey}. In the example of
entangled spinning particles the states of the parts are represented
by reduced density matrices and they are just a statistical mixture of
``up'' components and ``down'' components while the complete entangled state
has more information, in particular information about correlation
between the events observed in each component.  For instance the spin
measurements of both particles will be correlated.  Whenever both
observers measure the spin in the same direction their results would
be opposite, ---up for Alice and down for Bob or vice versa---. The
observers could never have figured out these correlations by looking
at the individual systems in isolation without comparing their
measurements. The complete system has certain non locality such that
when one electron chose to answer up, the other necessarily needs to
chose down. Such correlation does not involve time. As it is well
known Bell's theorem establishes that it is not possible to explain
this kind of behavior assuming that each part follows a
pre-established set of instructions, in other words, assuming that
each part has some local hidden information telling it how to act
before each measurement. In other terms the state of the entangled
pair of particles given for instance by $\vert s=1,s_z=0\rangle$ is not
mathematically determined by the the states of the parts. In fact
there are two states $\vert s=1,s_z=0\rangle$ and $\vert s=0,s_z=0\rangle$ whose
parts are in identical states.

These correlation are generic for any entangled state. Any system in
an entangled state presents downward causation. The events produced by
different components of the system have correlations that do not
result from the states of the parts. In fact, the very
characterization of the entanglement of a pure state in terms of the
Von Neumann entropy or entanglement entropy that is always a positive
quantity shows that the complete information required for the
determination of the whole cannot be recovered from the information of
the parts or that the state of the whole cannot be mathematically
expressed in terms of the states of its parts. The states' roles in
causation, together with their non-separability when they are
entangled, imply downward causation.

Less trivial explicit examples of downward causation may be found in
the molecular behavior where entanglement cannot be ignored or in
quantum computers.  In quantum chemistry calculations, entanglement is
related with the correlation energy. This energy is neglected in
Hartree--Fock calculations and the energy error of the Hartree--Fock
approximation to the wavefunctions of a molecular system is a measure
of the effects of downward causation in the behavior of the molecular
components.  In fact, the Hartre-Fock approximation consists in
writing the state of the electronic system as a tensor product of one
particle states. The approximation to the exact energy would be worse
when the system becomes more entangled.  For instance, it is downward
causation of the whole molecular system state that determines the
precise vibrational behavior of the nuclei.
  
For quantum computers, the existence of quantum correlations in the
entangled states between different input and output outcomes is at the
basis of the application of quantum algorithm that allow to solve
certain problems like integer factorization using Shor's algorithm
much more quickly than any classical computers.  The disposition of
the quantum computer to produce the correct correlation between the
input and output results at the end of the computation is the
manifestation of the quantum downward causation that is at the basis
of the improvement of the computational capabilities of quantum
computer.

We have characterized the downward causation of a system by its
disposition to produce certain effects that is not present in the
dispositions of its parts. This kind of disposition is the
characteristic feature of systems in entangled states. In turn, the
existence of entangled states in quantum systems result from the
exponential growth of Hilbert spaces of composed systems in opposition
to the linear growth of the states in classical physics. At the basis
of the novelty and non-reducibility of emergent systems it is this
exponential growth in the possible behaviors of the quantum systems
whose philosophical implications can be recognized when the
appropriate ontology is put into action.  Summarizing, emergence is
not the exception but the rule in interacting quantum systems, its so
natural that it usually remains unnoticed.  Strong emergence manifest
itself in most complex systems, it is a natural result of the
interaction of quantum objects.

The recent advances in the understanding of the role of quantum
mechanics in biology \cite{libro} and the previous analysis of strong
emergence in quantum mechanics rise the expectations of understanding
mental phenomena and their causal powers in physical terms. The issue
of emergence, also known as non reductive physicalism in the context
of studies of the mind brain problem, has been extensively analyzed
mostly using notions of supervenience that assume some form of
separability and are not valid in quantum mechanics as shown by
Maudlin \cite{maudlin}.  We consider that in quantum entangled systems with
downward causation and dispositional states that leads to
probabilistic outcomes the issue of free will may be posed in a clear
explicit way and will be analyzed elsewhere.

\section{Summary}

Several interpretations of quantum mechanics admit event ontologies.
These realistic interpretations lead to an important revision of the
notion of matter and its potentialities.  Systems of particles in
entangled states have new behaviors and emergent properties.

The quantum theory implies that the lower levels are modified even up
to the point where they lose part of their individuality when they
integrate into an entangled system in a higher level of the
hierarchy. The emergent structure has novel properties and downward
causation. Interpretations of quantum mechanics that admit an event ontology
solve the traditional problem of explaining emergence.

This work was supported in part by grant NSF-PHY-1305000,
NSF-PHY-1603063, ANII
FCE-1-2014-1-103974, funds of the Hearne Institute for Theoretical
Physics, CCT-LSU and Pedeciba.

\end{document}